\documentclass[a4paper,11pt]{article}

\usepackage{jcappub} 

\usepackage[T1]{fontenc} 
\usepackage{graphicx}
\usepackage{dcolumn}
\usepackage{amsfonts}
\usepackage{amssymb}
\usepackage{amsmath}
\usepackage{graphicx}
\usepackage{subfigure}
\usepackage{makeidx}
\usepackage{graphicx}
\usepackage{capt-of}
\usepackage{float}
\usepackage{placeins}
\usepackage{perpage}
\graphicspath{{images/}}

\title{Hamilton-Jacobi formalism for inflation with non-minimal derivative coupling}

\author[a]{Haidar Sheikhahmadi
}
\author[b, c]{Emmanuel N. Saridakis}
\author[d]{Ali Aghamohammadi}
\author[e]{Khaled Saaidi}

\affiliation[a]{Institute for Advance Studies in Basic Sciences (IASBS) Gava Zang, Zanjan
45137-
66731, Iran}
\affiliation[b]{Instituto de F\'{\i}sica, Pontificia Universidad de Cat\'olica de
Valpara\'{\i}so,
Casilla 4950, Valpara\'{\i}so, Chile}
\affiliation[c]{CASPER, Physics Department, Baylor University, Waco, TX  76798-7310, USA}
\affiliation[d]{ Sanandaj Branch Islamic Azad University,  Iran}
\affiliation[e]{Department of Physics, Faculty of Science, University of Kurdistan,
Sanandaj, Iran}
\emailAdd{h.sh.ahmadi@gmail.com}
\emailAdd{Emmanuel$_-$Saridakis@baylor.edu}
\emailAdd{a.aqamohamadi@iausdj.ac.ir}
\emailAdd{ksaaidi@uok.ac.ir}

\abstract{In inflation  with nonminimal derivative coupling there is not a conformal
transformation to the Einstein frame where calculations are straightforward, and thus in
order to extract inflationary observables one needs to perform a detailed and lengthy
perturbation investigation. In this work we bypass this problem by performing
a Hamilton-Jacobi analysis, namely rewriting the cosmological equations
considering the scalar field to be the time variable. We apply the method to two
specific models, namely the power-law and the exponential cases, and for each model we
calculate various observables such as the tensor-to-scalar ratio, and the spectral index
and its running. We compare them with 2013 and 2015 Planck data, and we show that they
are in a very good agreement with observations.}

\begin{document}
\maketitle
\flushbottom

\section{Introduction}
\label{Introduction}

After more than three decades of extensive investigation, the inflationary paradigm is
considered to be a necessary part of the Standard Model of cosmology, solving some of its
earlier crucial problems, such as the flatness, the horizon and the monopole ones
\cite{Kazanas:1980tx,Guth:1980zm,Starobinsky:1980te,Sato:1980yn,Starobinsky:1982ee,
Linde:1981mu}.
Additionally, inflation is needed in order to obtain the correct behavior of primordial
fluctuations and a universe with a nearly scale-invariant density power
spectrum
\cite{Liddle:1999mq,Langlois:2004de,Lyth:1998xn,Guth:2000ka,Lidsey:1995np,Bassett:2005xm},
 as well as with the correct amount of tensor perturbations
\cite{Grishchuk:1974ny,Starobinsky:1979ty,Allen:1987bk,Sahni:1990tx,Souradeep:1992sm,
Giovannini:1999qj,Sami:2004xk,Hossain:2014zma,Geng:2015fla,Hossain:2014coa,Cai:2014uka}.

In principle there are two main ways that one can obtain the realization of the
inflationary paradigm. The first direction is to use a modification of the gravitational
sector \cite{Nojiri:2003ft,Nojiri:2005pu,Capozziello:2005tf} (for a review see
\cite{Capozziello:2011et}), acquiring a modified cosmological
behavior that allows for inflationary solutions. The most well known scenario in this
approach is the Starobinsky inflation \cite{Starobinsky:1980te}, in which one adds in the
Einstein-Hilbert action a term quadratic in the Ricci scalar. The second direction is to
introduce new, exotic forms of matter, capable of driving inflation even in the framework
of general relativity. In this approach one usually adds a canonical scalar field,
assuming it to take large values (for instance in chaotic inflation \cite{Linde:1983gd})
or small values (for instance in new and natural inflation
\cite{Albrecht:1982wi,Freese:1990rb}), a phantom field
\cite{Piao:2004tq,Lidsey:2004xd,Elizalde:2008yf,Feng:2010ya}, a tachyon field
\cite{Fairbairn:2002yp,Feinstein:2002aj,Aghamohammadi:2014aca}, or
scenarios like k-inflation \cite{ArmendarizPicon:1999rj,Sheikhahmadi:2015gaa} and ghost
inflation \cite{ArkaniHamed:2003uz}.

Apart from the above simple inflationary realizations, one could construct models of both
contributions, namely models where the extra scalar field couples to gravity in a more
complicated way than the usual minimal coupling. The simplest class of such scenarios is
when the scalar field is non-minimally coupled to gravity, and indeed these
``scalar-tensor'' theories present very interesting cosmological behavior
\cite{Fakir:1990eg,Sahni:1998at,Uzan:1999ch,
Faraoni:2000wk,Tsujikawa:2000tm,Piao:2002nh,Tsujikawa:2004my,Kaiser:2013sna}. However, an
interesting class of models is obtained if one extends further these
constructions by allowing for non-minimal couplings between the
curvature and the derivatives of the scalar field \cite{Amendola:1993uh}, which can lead
to novel and interesting cosmological features
\cite{Capozziello:1999uwa,Capozziello:1999xt,Daniel:2007kk,Balakin:2008cx,
Saridakis:2010mf,Gao:2010vr,
Germani:2010gm,
Granda:2011zk,Shchigolev:2011nma,
Tsujikawa:2012mk,Sadjadi:2012zp,Sadjadi:2013na,
Koutsoumbas:2013boa,
Feng:2013pba,Sadjadi:2013psa,Dent:2013awa,Skugoreva:2014gka,Yang:2015pga,
Koutsoumbas:2015ekk}. Finally, even more complicated extensions can arise considering
Galileon and Horndeski theories \cite{Horndeski:1974wa,Nicolis:2008in,Deffayet:2009wt} or
even bi-scalar and multi-scalar constructions
\cite{Padilla:2010tj,Saridakis:2016ahq,Saridakis:2016mjd}.

In the most studies of inflationary cosmology one imposes the usual slow-roll
approximation, and tries to extract expressions for basic inflationary-related
observables, such as the scalar and tensor spectral indices, the running spectral index,
and the tensor-to-scalar
ratio. Nevertheless, there is an alternative approach which allows for an easier
derivation of many inflation results, namely the Hamilton-Jacobi formulation
\cite{Salopek:1990jq}. In this formalism, one rewrites the cosmological equations by
considering the scalar field to be the time variable, which is always possible during the
slow-roll era, where the scalar varies monotonically (the extension to the
post-inflationary, oscillatory epoch is straightforward by matching together separate
monotonic epochs \cite{Liddle:2000cg}). We mention here that in the usual approach it is
more convenient to transform to the Einstein frame, where calculations are significantly
easier, and hence this approach cannot be easily applied to models where there is not a
conformal transformation to such a frame, such as the nonminimal derivative coupling
constructions. Hence, in such cases we expect the Hamilton-Jacobi formalism to be more
convenient and significantly easier.

In this work we are interested in performing the Hamilton-Jacobi analysis for inflation
with nonminimal derivative couplings. The plan of the work is as follows:
In Section \ref{model} we briefly review inflation with nonminimal derivative coupling
and then we apply the Hamilton-Jacobi formulation. Then in Section \ref{TYINS} we apply
it to two specific models, namely the power-law and the exponential cases. We calculate
various observables such as the tensor-to-scalar ratio, and the spectral index and its
running, and we compare them with the Planck data. Finally, Section \ref{Conc} is
devoted to final remarks and conclusion.

\section{Hamilton-Jacobi formalism for inflation with nonminimal derivative coupling}
\label{model}

In this section we will construct the Hamilton-Jacobi formalism for inflation with
nonminimal derivative coupling. We first give a brief review of cosmology with nonminimal
derivative couplings, and then we proceed to the Hamilton-Jacobi formulation.

\subsection{Inflation with nonminimal derivative coupling}

Let us start with
the inflation realization in cosmology with nonminimal derivative couplings. The action
of such a theory reads as  \cite{Amendola:1993uh,Saridakis:2010mf}
\begin{eqnarray}
\label{action1}
&&\!\!\!\!\!\!\!\!
S_{\phi}=\int d^4x\sqrt{-g}\left[{M_P^2\over 2}R-{1\over 2}g^{\mu
\nu}\partial_\mu \phi\partial_\nu \phi+{1\over 2M^2}G^{\mu
\nu}\partial_\mu \phi \partial_\nu \phi -V(\phi)\right],
\end{eqnarray}
where $G^{\mu \nu}=R^{\mu \nu}-{1\over 2}g^{\mu \nu}R$ is the
Einstein tensor, $R$ the scalar Ricci,    $M_P=\sqrt{{1\over 8\pi G}}$
the
reduced Planck mass, and $M$ is the coupling constant with  dimension of mass. Since in
this work we focus on the inflationary application of this theory, we have neglected the
matter and radiation contents.
Variation of the above action in terms of the
metric gives rise to the field equations
\begin{equation}
\label{eqn0} G_{\mu\nu}=\frac{1}{M_P^2}\left[
T^{(\phi)}_{\mu\nu}
+\frac{1}{M^2}\Theta_{\mu\nu}\right]-\frac{1}{M_P^2} g_{\mu\nu} V(\phi),
\end{equation}
with
\begin{eqnarray} &&\!\!\!\!\!\!\!\!\!\!\!\!\!\!\!\!\!\!\!\!
T^{(\phi)}_{\mu\nu}=\nabla_\mu\phi\nabla_\nu\phi-
{\textstyle\frac12}g_{\mu\nu}(\nabla\phi)^2,
\nonumber\\
&&\!\!\!\!\!\!\!\!\!\!\!\!\!\!\!\!\!\!\!
\Theta_{\mu\nu}=-{\textstyle\frac12}\nabla_\mu\phi\,\nabla_\nu\phi\,R
+2\nabla_\alpha\phi\,\nabla_{(\mu}\phi R^\alpha_{\nu)}
+\nabla^\alpha\phi\,\nabla^\beta\phi\,R_{\mu\alpha\nu\beta}
+\nabla_\mu\nabla^\alpha\phi\,\nabla_\nu\nabla_\alpha\phi
\nonumber\\
&&
-\nabla_\mu\nabla_\nu\phi\,\square\phi-{\textstyle\frac12}(\nabla\phi)^2
G_{\mu\nu}+g_{\mu\nu}\big[-{\textstyle\frac12}\nabla^\alpha\nabla^\beta\phi\,
\nabla_\alpha\nabla_\beta\phi+{\textstyle\frac12}(\square\phi)^2
-\nabla_\alpha\phi\,\nabla_\beta\phi\,R^{\alpha\beta}
\big], \nonumber
\end{eqnarray}
where  $\nabla_{(\mu}\phi R^{\alpha}_{\nu)} = \frac{1}{2}(\nabla_{\mu}\phi
R^{\alpha}_{\nu}+\nabla_{\nu}\phi R^{\alpha}_{\mu})
$.
Additionally, variation of the action
(\ref{action1}) with respect to $\phi$ provides the scalar field
equation of motion, namely
\begin{eqnarray}
\label{eqnphi0}
 \left[
g^{\mu\nu}+\frac{1}{M^2} G^{\mu\nu}\right]\nabla_{\mu}\nabla_\nu\phi=V'(\phi),
\end{eqnarray}
 where
$V'(\phi)\equiv dV(\phi)/d\phi$.

In order to investigate the cosmological applications of
the above theory, we focus on a spatially-flat
Friedmann-Robertson-Walker (FRW) background geometry of the form
\begin{equation}
\label{FRW0metric0}
ds^2= -dt^2+a^2(t)\,\delta_{ij} dx^i dx^j,
\end{equation}
where $t$ is the cosmic time, $x^i$ are the comoving spatial coordinates and
$a(t)$ is the scale factor. In this case the field equations (\ref{eqn0}) give rise to
the two Friedmann equations, namely
\begin{eqnarray}
\label{FR1}
&&H^2={1\over 3M_P^2}\rho_\phi\\
&&\dot{H}=-{1\over 2M_P^2}\left(\rho_\phi+P_\phi\right),
\end{eqnarray}
 with $H=\dot{a}/a$ the Hubble parameter (a
dot denotes differentiation with respect to $t$), and where we have introduced the
effective energy density and pressure of the scalar field respectively as
\begin{eqnarray}\label{rho3}
&&\!\!\!\!\!\!\!\!\!\!
\rho_{\phi}={1\over 2}\left(1+9{H^2\over
M^2}\right)\dot{\phi}^2+V(\phi)\\
&&\!\!\!\!\!\!\!\!\!\!
P_{\phi}=\frac{\dot{\phi}^2}{2}
\left[1-\frac{1}{M^2}\left(2\dot{H}+3H^2+\frac{4H
\ddot{\phi}}{\dot{\phi}}\right)\right]
-V(\phi).
\end{eqnarray}
Similarly, the scalar-field equation of motion (\ref{eqnphi0}) becomes
\begin{equation}
\label{eqnphi5}
\left(1+{3H^2\over M^2}\right)\ddot{\phi}+3H\left(1+{3H^2\over
M^2}+{{2\dot{H}}\over M^2}\right)\dot{\phi}+V'(\phi)=0.
\end{equation}
Note that using the definitions of $\rho_{\phi}$ and $P_\phi$ we can re-write this
equation in the usual conservation form, namely
\begin{equation}
\label{conserv6}
\dot{\rho_{\phi}}+3H(P_{\phi}+\rho_{\phi})=0.
\end{equation}
Finally, we stress here that, as it is well known, the
equations of motion  do not contain higher-order time derivatives,
and thus the theory at hand is ghost free \cite{Amendola:1993uh,Saridakis:2010mf}.

Lastly, since in this work we focus on the inflation realization, we restrict ourselves
to the high friction regime \cite{Germani:2010gm,Sadjadi:2012zp} where ${H^2/M^2}\gg1$,
and we impose the slow-roll conditions, namely ${H^2\over M^2}\dot{\phi}^2\ll V(\phi)$,
$\ddot{\phi}\ll H\dot{\phi}$. Hence, the first Friedmann equation (\ref{FR1}) and the
scalar-field evolution equation (\ref{eqnphi5}) respectively become
\begin{eqnarray}
\label{8a}
&&H^2\approx {1\over 3M_P^2}V(\phi) \\
&&\dot{\phi}\approx -{M^2V'(\phi)\over 9H^3}\,.
\label{8b}
\end{eqnarray}

\subsection{Hamilton-Jacobi formalism}

Let us now formulate the Hamilton-Jacobi approach to inflation with nonminimal
derivative couplings. We first describe briefly the main idea of Hamilton-Jacobi
formalism \cite{Salopek:1990jq}. In this approach of a cosmological system, one uses the
scalar field as a time variable, and hence the Friedmann equation gives rise to a partial
differential equation for the Hubble parameter. Thus, concerning inflation, one imposes
the slow-roll conditions, and then by choosing suitable ansatzes he can extract
analytical
solutions, as well as explicit expressions for the inflationary observables. Note that
the Hamilton-Jacobi formalism is very efficient since it can bypass the extensive
calculations that are needed in the usual approach, especially in the case where the
transformation to the Einstein frame (where calculations are easier) is impossible, such
is the case of cosmology with nonminimal derivative couplings. We mention that in order
for this procedure to be self-determined, we need a monotonically varying scalar field,
which is indeed the case during the slow-roll era. Even for the post-inflationary case,
where the scalar field is expected to oscillate, one can still apply the Hamilton-Jacobi
formalism, by matching together separate monotonic epochs \cite{Liddle:2000cg}.

We now apply the above into the slow-roll inflationary cosmological equations
(\ref{8a}) and (\ref{8b}). First of all, we can combine them in order to
obtain the useful relation
\begin{equation}
\label{eq9}
\dot{\phi}\approx -\frac{2}{3}{M^2 M_P^2 H'(\phi)\over H^2}\, .
\end{equation}
As we observe, it is obvious that if $H'(\phi)<0~(H'(\phi)>0)$ then the scalar field
increases (decreases) over time.
Now, inserting (\ref{eq9}) into the first Friedmann equation we are led to the
Hamilton-Jacobi
equation, namely
\begin{eqnarray}
\label{eq10}
&&[H'(\phi)]^{2}-\frac{3}{2}\frac{H^{6}(\phi)}{M^{2}[9H^{2}(\phi)+M^2]}+\frac{
V(\phi)H^ { 4 }
(\phi)}{
18 M_P^2 M^{2}[9H^{2}(\phi)+M^2]}=0\, .
\end{eqnarray}
Thence, we can express the potential in terms of the scalar field as
\begin{eqnarray}
\label{eq11}
&&\!\!\!\!\!\!\!\!
V(\phi)=27 M_P^2H^{2}(\phi)-\frac{{18}  M_P^2 M^{2}
[H'(\phi)]^{2}[9H^{2}(\phi)+M^2]}{
H^{4}(\phi)}\, .
\end{eqnarray}

We now introduce the slow-roll parameters \cite{Martin:2013tda}, which using (\ref{eq9})
they finally become:
\begin{eqnarray}
\label{eq12}
&&\epsilon_1(\phi)
\equiv-\frac{\dot{H}}{H^2}\approx
\frac{2}{3}M^2  M_P^2
\left[\frac{H'(\phi)}{H^{2}(\phi)}\right]^{2}\, ,
\\
&&
\epsilon_2(\phi)
\equiv  \frac{\ddot{H}}{H\dot{H}}-\frac{2\dot{H}}{H^2}\approx
-\frac{4}{3} M^2  M_P^2
\left\{\frac{H''(\phi)}{H^{3}(\phi)}
-2
\left[\frac{H'(\phi)}{H^{2}(\phi)}\right]^{2}
\right\}
 \, ,
\label{eq12b}
\end{eqnarray}
which as usual are both much smaller than unity during slow-roll inflation. Moreover, the
time when $\epsilon_1$ becomes equal to one marks the end of inflation.

Combining relations (\ref{eq9}) and $\dot{a}=\dot{\phi}{a}^\prime$, we can extract the
scale factor as
\begin{eqnarray}
\label{eq14}
a(t)=a_0\,\exp\left[{-\frac{3}{2}\int M_P^{-2}
M^{-2}\frac{H^{3}(\phi)}{H'(\phi)}d\phi}\right]\, ,
\end{eqnarray}
where $a_0$ is an integration constant. Thus, the e-folding number, which describes the
amount of expansion during the inflation, can be calculated as
\begin{eqnarray}
\label{eq15}
N\equiv\int_{t_{i}}^{t_{e}} H(t)dt=\int_{\phi_{i}}^{\phi_{e}}
\frac{H(\phi)}{\dot{\phi}}d\phi\, ,
\end{eqnarray}
where the subscripts ``i'' and ``e''  denote the initiation and end of inflation.

\section{Applications}
\label{TYINS}

In the previous section we presented the Hamilton-Jacobi formulation of inflation with
nonminimal derivative couplings. Hence, in this section we can proceed to the
investigation of specific applications, considering specific ansatzes for the Hubble
function. In particular, in the following subsections we consider the power-law and the
exponential cases separately.

\subsection{Hubble parameter as power-law function}
\label{POWELLAW}

Let us first examine the case where the Hubble parameter is a power-law function of the
scalar field, namely
\begin{equation}
H(\phi)=\alpha \phi^n,
\label{powansatz}
\end{equation}
 where $\alpha$ and $n$ are the model
parameters. In this case  Eq. (\ref{eq9}) becomes
\begin{eqnarray}
\label{eq18}
&&\dot{\phi}\approx -\frac{2}{3}{M^2 M_P^2 n\over \alpha \phi^{n+1}}\, ,
\end{eqnarray}
and thus substituting into (\ref{eq11}) we acquire the potential as
\begin{eqnarray}
\label{eq19}
&&\!\!\!\!\!\!\!
V(\phi)=27 M_P^2\alpha^{2}\phi^{2n}
-18  M_P^2 M^{2}
\frac{n^2}{\alpha^{2}\phi^{2(n+1)}}\left[9\alpha^{2}\phi^{2n}+M^2\right]\, .
\end{eqnarray}
The above potential is a sum of power and inverse power laws, and, although slightly
complicated, potentials of these forms are often used in cosmological applications
\cite{Peebles:1987ek,Abramo:2003cp,Aguirregabiria:2004xd,
Copeland:2004hq,Saridakis:2009pj,Skugoreva:2013ooa}.

Additionally, the slow-roll parameters (\ref{eq12}),(\ref{eq12b}) respectively become
\begin{eqnarray}
\label{eq22}
&&\epsilon_1(\phi)\approx
\frac{2n^2}{3\alpha^2}M^2M_P^2 \, \phi^{-2(n+1)}\, ,\\
&&
\epsilon_2(\phi)
\approx
\frac{4n(n+1)}{3\alpha^2} M^2M_P^2 \,
\phi^{-2(n+1)}\, .
\label{eq22b}
\end{eqnarray}
Therefore, imposing $\epsilon_1(\phi)=1$  provides the scalar field
at the end of inflation as
\begin{eqnarray}
\label{eq20}
\phi_{e}=   \left(\frac{2M^2M_P^2n^2}{3\alpha^2}\right)^{\frac{1}{2(n+1)}}
 \, ,
\end{eqnarray}
while using (\ref{eq15}) and (\ref{powansatz}),(\ref{eq18}) we can calculate the value of
the scalar field at the beginning
of inflation in terms of the e-folding number as
\begin{equation}
\label{eq21}
\phi_{i}=
 \left(\frac{2 M^2M_P^2\,n}{3\alpha^2}\right)^{\frac{1}{2(n+1)}}\left[
2(n+1) N   +n\right]^{\frac{1}{2(n+1)}}.
\end{equation}

We can now use the above expressions in order to calculate the inflationary-related
observables, namely the scalar and tensor spectral indices, and the tensor-to-scalar
ratio respectively as  \cite{Martin:2013tda}
\begin{eqnarray}
\label{eq23}
n_{S}-1&\simeq&
-2\epsilon_1(\phi_{i})-2\epsilon_2(\phi_{i})=
-\frac{2(3n+1)}{2(n+1)N+n }  \, ,\\
\label{eq24}
n_{T}&\simeq&-2\epsilon_1(\phi_{i})=-\frac{2n}{2(n+1)N+n } \, ,\\
r&=&-8n_T\, .
\label{eq23bb}
\end{eqnarray}
Hence, eliminating $n$ between (\ref{eq23}),(\ref{eq24}) and (\ref{eq23bb}) we can obtain
\begin{equation}
\label{rnspowerlaw}
r=\frac{16 (N-1)}{4N-1}-\frac{16 N}{4N-1}n_S,
\end{equation}
and
\begin{equation}
\label{rnTpowerlaw}
n_T=-\frac{2 (N-1)}{4N-1}+\frac{2 N}{4N-1}n_S,
\end{equation}
which prove to be very useful, since they allow us to confront our model predictions
straightaway with the data. Note that the parameter $n$ does not appear
in the final parametric expressions, which is very good since it decreases the number of
free fitting parameters.
\begin{figure}[ht]
\centering
\includegraphics[scale=.50]{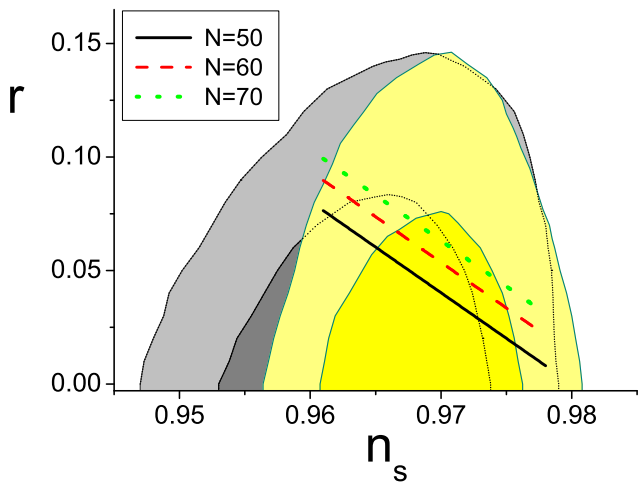}
\caption{{\it{ 1$\sigma$ (yellow) and 2$\sigma$ (light yellow) contours for Planck 2015
results ($TT+lowP+lensing+BAO+JLA+H_0$)  \cite{Ade:2015xua}, and 1$\sigma$ (grey) and
2$\sigma$ (light grey) contours for Planck 2013 results  ($Planck+WP+BAO$)
\cite{Planck:2013jfk} (note that the  1$\sigma$ region of Planck 2013 results  is behind
the Planck 2015 results, hence we mark its boundary by a dotted curve), on $n_S-r$ plane.
Additionally, we depict the predictions of our scenario, for the power-law case
(\ref{powansatz}),  with the e-folding value
$N$ being $50$, $60$ and $70$.
}}}
\label{powerrns}
\end{figure}

Finally, the running of the scalar spectral
index is defined as
\begin{equation}
\label{runningTpowerlaw}
\alpha_{S}\equiv\frac{d n_{S}}{d \ln k}.
\end{equation}
Hence, since from the definitions
 of $n_S$ and $n_T$  one can find \cite{Baumann}
\begin{equation}
\label{runningTpowerlaw1}
\frac{d N}{d \ln k}=\left(1+\frac{d \ln H}{d \ln k}\right)^{-1}\simeq(1+\epsilon_1),
\end{equation}
we can use  (\ref{eq23}), (\ref{runningTpowerlaw}) and
(\ref{runningTpowerlaw1}) in order to express $\alpha_{S}$ as
\begin{eqnarray}
\label{runningTpowerlaw2}
&&\!\!\!\!\!\!\!\!\!\!\!\!\!\!\!
\alpha_{S}=\frac{d n_{S}}{d N} \frac{d N}{d \ln k}\simeq
\frac{8(n+1)(3n+1)[(n+1)N+n]}{[2(n+1)N+n]^3}.
\end{eqnarray}
\begin{figure}[ht]
\centering
\includegraphics[scale=.47]{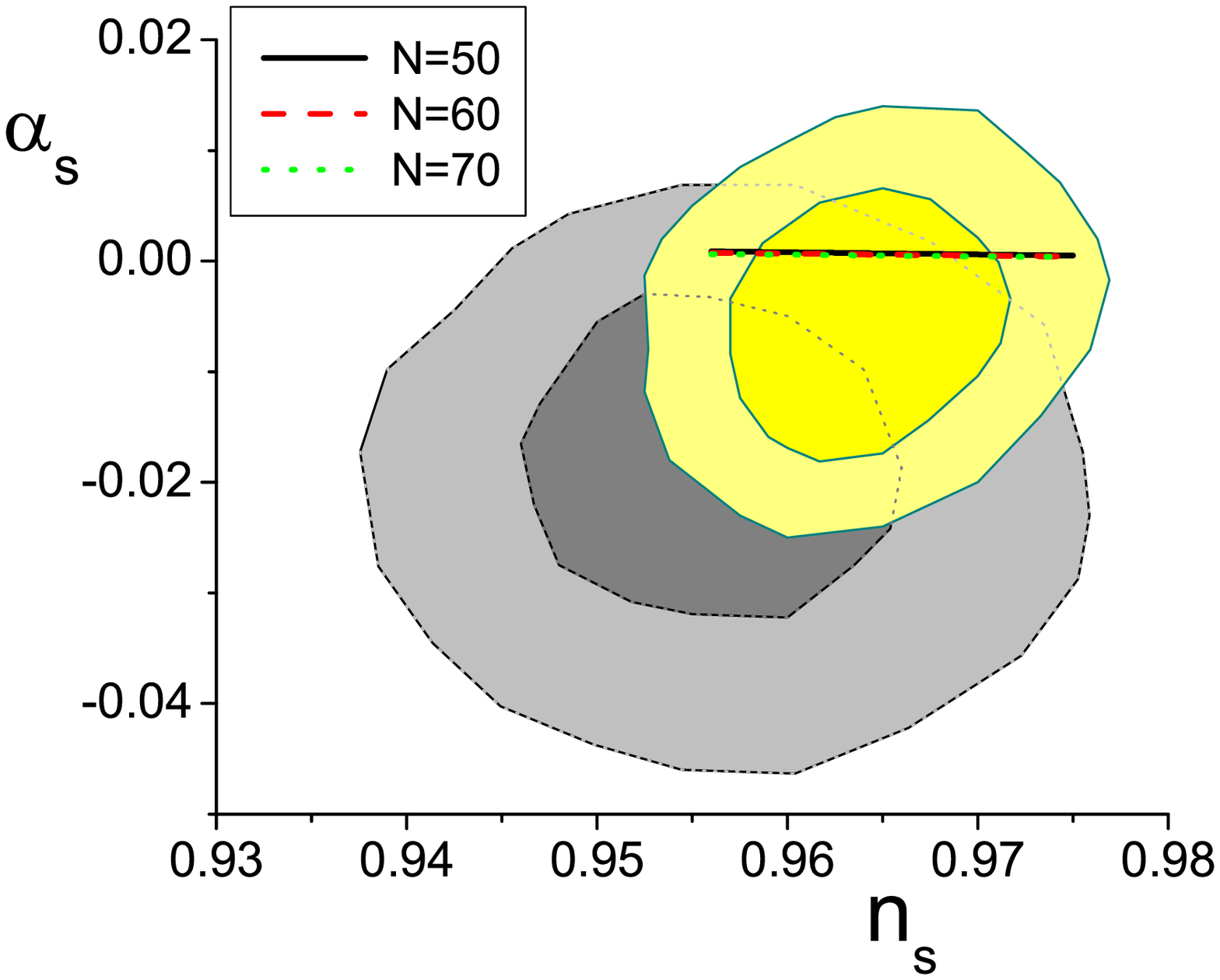}
\caption{{\it{ 1$\sigma$ (yellow) and 2$\sigma$ (light yellow) contours for Planck 2015
results ($TT,TE,EE+lowP$)  \cite{Ade:2015xua}, and 1$\sigma$ (grey) and
2$\sigma$ (light grey) contours for   Planck 2013 results  ($\Lambda
CDM+running+tensors$)
\cite{Planck:2013jfk}, on $n_S-\alpha_S$ plane. Additionally, we depict the predictions of
our scenario, for    the power-law case
(\ref{powansatz}),  with the e-folding value
$N$ being $50$, $60$ and $70$. The various  curves are indistinguishable in the
resolution scale of the graph.  }}}
\label{runningpow}
\end{figure}
Thus, eliminating $n$ between  (\ref{eq23}) and (\ref{runningTpowerlaw2}) we obtain
$\alpha_S$ as a function of $n_S$, namely
\begin{equation}
\label{runningTpowerlaw2bb}
\alpha_{S}\simeq
\frac{(n_S-1)(n_S+3)[(n_S-5)N+2]}{(4N-1)^2}.
\end{equation}
\begin{figure}[ht]
\centering
\includegraphics[scale=.42]{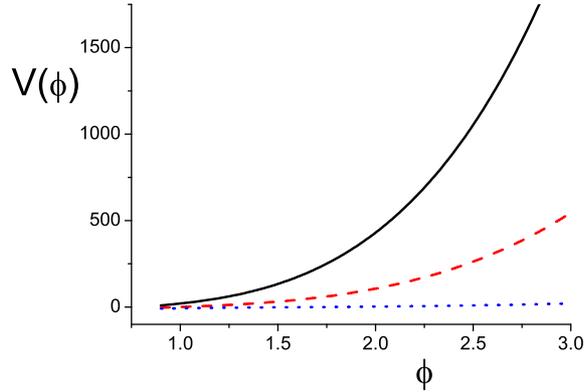}
\caption{{\it{ The scalar potential $V(\phi)$ according to expression (\ref{eq19}), for
the power-law case
(\ref{powansatz}), for $M=0.1$,  $\alpha=1$, $n=2$ (black-solid), for $M=0.1$,
$\alpha=0.5$, $n=2$ (red-dashed), and for $M=0.1$,  $\alpha=0.1$, $n=2$ (blue-dotted).
All quantities are measured in $M_P$ units.}}}
\label{powerpotential}
\end{figure}

In order to present these features more transparently, in
Fig.~\ref{powerrns} we present the predictions of our scenario with the e-folding value
${N}$ being $50$, $60$ and $70$, on top of the 1$\sigma$ and 2$\sigma$ contours of
the Planck 2013 results \cite{Planck:2013jfk} as well as of the  Planck 2015 results
\cite{Ade:2015xua}. As we observe, the scenario at hand is in very good agreement with
observations, with the agreement being better for lower $N$.
Additionally, in Fig. \ref{runningpow} we depict the predictions of the scenario
at hand for the running    spectral index $\alpha_{S}$
with the e-folding value ${N}$
being $50$, $60$ and $70$,  on top of the 1$\sigma$ and 2$\sigma$ contours of the Planck
2013
results \cite{Planck:2013jfk} as well as of the  Planck 2015 results
\cite{Ade:2015xua}. The  agreement
with observations is very satisfactory.

 Finally, in order to provide a picture of the scalar potential, we use
expression (\ref{eq19}) and  in Fig. \ref{powerpotential}  we depict $V(\phi)$ for
various values of $M$, $\alpha$ and $n$.  In this Figure we have focused on the
$\phi$-values in which $\phi_{i}$ according to
(\ref{eq21}) and $\phi_{e}$ according to (\ref{eq20}) lie.

\subsection{Hubble parameter as exponential function}
\label{EXPONENTIAL}

In this subsection we study the case where the Hubble parameter is an exponential function
of the scalar field, namely
\begin{equation}
H(\phi)=\beta e^{q\phi},
\label{expansatz1}
\end{equation}
 where $\beta$ and $q$ are the
model parameters. In this case  Eq. (\ref{eq9}) gives
\begin{eqnarray}
\label{eq27}
&&\dot{\phi}\approx -\frac{2q}{3\beta}{M^2 M_P^2 e^{-q\phi}}\,,
\end{eqnarray}
and therefore inserting this expression into (\ref{eq11}) we acquire the potential as
\begin{eqnarray}
\label{eq28}
&&\!\!\!\!\!\!\!
V(\phi)=27 M_P^2\beta^{2}e^{2q\phi}-18 M_P^2 M^{2}
\frac{q^2}{\beta^{2}e^{2q\phi}}\left(9\beta^{2}
e^{2q\phi}+M^2\right)\, .
\end{eqnarray}
The above potential is a sum of exponential potentials, and potentials of these forms are
often used in cosmological applications
\cite{Halliwell:1986ja,Aguirregabiria:1993pm,Aguirregabiria:1993pk,Coley:1997nk,
Copeland:1997et,Barreiro:1999zs,Heard:2002dr,Rubano:2003et}.

Furthermore, the slow-roll parameters (\ref{eq12}),(\ref{eq12b})  become respectively
\begin{eqnarray}
\label{eq31}
&&\epsilon_1(\phi)\approx \frac{2q^2}{3\beta^2}M^2M_P^2\,e^{-2q\phi}\, ,\\
&&
\epsilon_2(\phi)\approx \frac{4q^2}{3\beta^2}M^2M_P^2\,e^{-2q\phi}\, ,
\label{eq31bbbb}
\end{eqnarray}
and we can then impose the condition $\epsilon_1(\phi)=1$ in order to calculate the scalar
field at the end of inflation as
\begin{eqnarray}
\label{eq29}
\phi_{e}=\frac{1}{2q}\ln\left(\frac{2M^2M_P^2\,q^2 }{3\beta^2}\right)\, .
\end{eqnarray}
Additionally, we can use (\ref{eq15}) and (\ref{expansatz1}),(\ref{eq27}) and
extract the value of the scalar field at the
beginning of inflation as a function of the e-folding number as
\begin{eqnarray}
\label{eq30}
\phi_{i}=\frac{1}{2q}\ln\left[ \frac{2 M^2M_P^2\,q^2}{3\beta^2}  \left(2 N+1\right)
\right]\, .
\end{eqnarray}

Concerning the scalar and tensor spectral indices and the tensor-to-scalar ratio, we
obtain \cite{Martin:2013tda}
\begin{eqnarray}
\label{eq32}
n_{S}-1&\simeq&
-2\epsilon_1(\phi_{i})-2\epsilon_2(\phi_{i})=
-\frac{6}{2N+1}\, ,\\
\label{eq33}
n_{T}&\simeq&-2\epsilon_1(\phi_{i})=-\frac{2}{2N+1}\,  ,\\
r&=&-8n_T\, ,
\label{eq33bb}
\end{eqnarray}
where from the first two of the above expressions we acquire
$n_{S}-1=3n_{T}$. Hence, from Eqs.(\ref{eq32}),(\ref{eq33}) and (\ref{eq33bb}) we can
obtain
\begin{equation}
\label{rnsexpon}
r=\frac{8}{3}-\frac{8}{3}n_{S}
\end{equation}
and
\begin{equation}
\label{rnTexpon}
n_T=-\frac{1}{3}+\frac{1}{3}n_{S},
\end{equation}
which prove to be very useful, since they allow us to confront our model predictions
straightaway with the data. Note that the parameters $\beta$ and $q$ do not appear in
the final parametric expressions, as well as the e-folding number $N$. Finally, the
running of the scalar spectral
index reads as
$
\alpha_{S}=\frac{d n_{S}}{d N} \frac{d N}{d \ln k} \simeq \frac{d n_{S}}{d N}\,
(1+\epsilon_1)$,
which with the help of (\ref{eq31}),(\ref{eq32}) becomes
\begin{eqnarray}
\label{runningTpowerlaw2bb}
&&\!\!\!\!\!\!\!\!\!\!\!\!\!\!\!
\alpha_{S}\simeq \frac{24(N+1)}{(2N+1)^3} ,
\end{eqnarray}
and therefore eliminating $N$ using (\ref{eq32}) we finally acquire
\begin{equation}
\label{runningTexpo2cc}
\alpha_{S}\simeq
\frac{1}{18}(n_{S}-7) (n_{S}-1)^2.
\end{equation}
 \begin{figure}[ht]
\centering
\includegraphics[scale=.50]{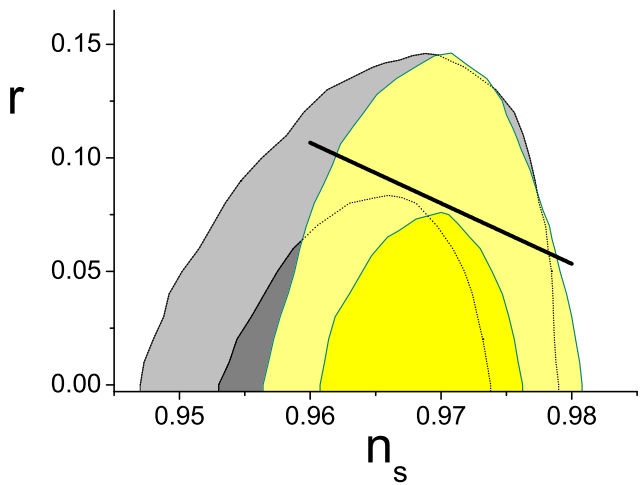}
\caption{{\it{ 1$\sigma$ (yellow) and 2$\sigma$ (light yellow) contours for Planck 2015
results ($TT+lowP+lensing+BAO+JLA+H_0$)  \cite{Ade:2015xua}, and 1$\sigma$ (grey) and
2$\sigma$ (light grey) contours for Planck 2013 results  ($Planck+WP+BAO$)
\cite{Planck:2013jfk} (note that the  1$\sigma$ region of Planck 2013 results  is behind
the Planck 2015 results, hence we mark its boundary by a dotted curve), on $n_S-r$ plane.
Additionally, with the black solid line we depict the prediction of our scenario, for
the exponential case (\ref{expansatz1}), which according to relation (\ref{rnsexpon})
does not depend on $\beta$, $q$ and $N$.
}}}
\label{exprns}
\end{figure}

\begin{figure}[ht]
\centering
\includegraphics[scale=.47]{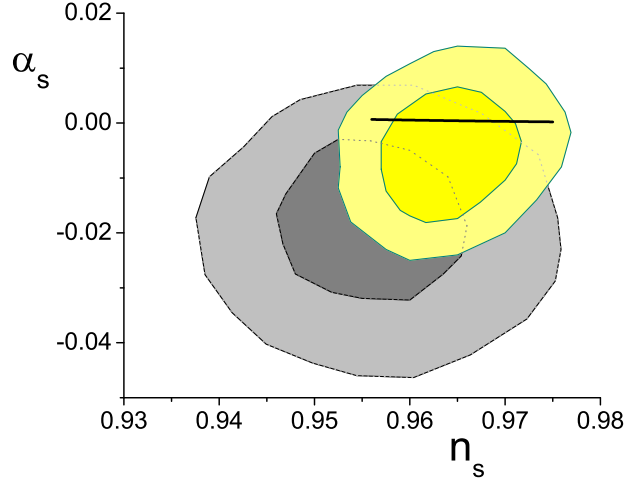}
\caption{{\it{ 1$\sigma$ (yellow) and 2$\sigma$ (light yellow) contours for Planck 2015
results ($TT,TE,EE+lowP$)  \cite{Ade:2015xua}, and 1$\sigma$ (grey) and
2$\sigma$ (light grey) contours for   Planck 2013 results  ($\Lambda
CDM+running+tensors$)
\cite{Planck:2013jfk}, on $n_S-\alpha_S$ plane. Additionally, we depict the predictions of
our scenario, for    the exponential case
(\ref{expansatz1}), for  the
e-folding value $N$ being $50$ and $60$. The two  curves are indistinguishable in the
resolution scale of the graph.  }}}
\label{runningexp}
\end{figure}

In order to present these features in a more clear way, in
Fig.~\ref{exprns} we illustrate the predictions of our scenario on $n_S-r$ plane,
 on top of the 1$\sigma$ and 2$\sigma$ contours of
the Planck 2013 results \cite{Planck:2013jfk} as well as of the  Planck 2015 results
\cite{Ade:2015xua}. As we mentioned above, according to relation (\ref{rnsexpon}) the
predictions of our scenario do not depend on
$\beta$, $q$ and $N$, however they are still in agreement with observations at 2$\sigma$
level. Furthermore,  in Fig. \ref{runningexp} we show the predictions of the scenario
at hand for the running    spectral index $\alpha_{S}$,  on top of the 1$\sigma$
and 2$\sigma$ contours of the Planck 2013 results \cite{Planck:2013jfk} as well as of the
Planck 2015 results \cite{Ade:2015xua}. According to relation (\ref{runningTexpo2cc}) the
model predictions are independent of $\beta$, $q$ and $N$, however they are in very good
agreement with observations.

\begin{figure}[ht]
\centering
\includegraphics[scale=.42]{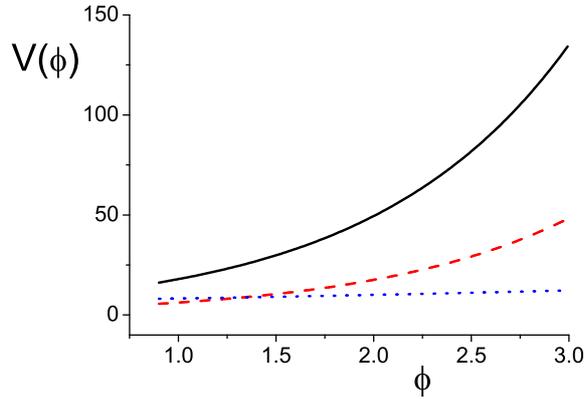}
\caption{{\it{ The scalar potential $V(\phi)$ according to expression (\ref{eq28}), for
the  exponential case
(\ref{expansatz1}), for $M=0.1$,  $\beta=0.5$, $q=0.5$ (black-solid), for $M=0.1$,
$\beta=0.3$, $q=0.5$ (red-dashed), and for $M=0.1$, $\beta=0.5$, $q=0.1$ (blue-dotted).
All quantities are measured in $M_P$ units.}}}
\label{expopotential}
\end{figure}

 Lastly, in order to give a picture of the scalar potential, we use
expression (\ref{eq28}) and  in Fig. \ref{expopotential}  we show $V(\phi)$ for
various values of $M$, $\beta$ and $q$.  In this Figure we have focused on the
$\phi$-values in which $\phi_{i}$ according to
(\ref{eq30}) and $\phi_{e}$ according to (\ref{eq29}) lie.

\subsection{Comparison with minimal-coupling case}

In this subsection we briefly compare the effect of the nonminimal derivative coupling
with the minimal-coupling case, for completeness (although in this simple case slow-roll
conditions are hard to be realized and moreover the unitarity bounds of the theory may
be violated \cite{Germani:2010gm,Burgess:2009ea,Lerner:2009na}).

Let us consider action (\ref{action1}) without the nonminimal derivative coupling term.
In this case, as it is well known, one obtains the standard Friedmann
equations, namely $3M_P^2{H^2} = \frac{{{\dot \phi }^2}}{2} + V(\phi )$ and
$3M_P^2(\dot{H}+{H^2}) = -  \frac{{{\dot \phi }^2}}{2} + V(\phi
)$, while the scalar-field equation reads $\ddot \phi  + 3H\dot \phi  + {V^{\prime}}(\phi)
= 0$. Hence, under the  slow-roll conditions,$
\frac{{{\dot \phi }^2}}{2} \ll V(\phi)$ and
$| {\ddot \phi } | \ll H | {\dot \phi } |$,   we obtain
 the following constraints
\begin{equation}
\label{Constraint1}
3 M_P^2 H^2\simeq V(\phi),
\end{equation}
\begin{equation}
\label{Constraint2}
{{\dot \phi }} \simeq -\frac{{V^{\prime}}(\phi)}{{3H}},
\end{equation}
which easily lead to  ${V^{\prime}}(\phi)\simeq  6 M_P^2 HH^{\prime}(\phi)$ and
thus to
\begin{equation}
\label{phidot}
\dot{\phi}\simeq- 2 M_P^2 H^{\prime}(\phi).
\end{equation}
Therefore, in this case, substitution into the first Friedmann equations gives the
Hamilton-Jacobi equation as
\begin{equation}
\label{H-J}
{H^{\prime}}^2(\phi)-  3  M_P^2 H^{2}(\phi)+\frac{ M_P^4}{2} V(\phi)=0,
\end{equation}
and thus the potential in terms of the scalar field is expressed as
\begin{equation}
\label{Potential1}
V(\phi)=\frac{6}{M_P^2}H^{2}(\phi)-\frac{2}{M_P^4}{H^{\prime}}^2(\phi).
\end{equation}
Finally,  the slow-roll parameters read as
\begin{eqnarray}
\label{eq12dd}
&&\epsilon_1(\phi)
\equiv-\frac{\dot{H}}{H^2}\approx
2  M_P^2
\left[\frac{H'(\phi)}{H(\phi)}\right]^{2}\, ,
\\
&&
\epsilon_2(\phi)
\equiv  \frac{\ddot{H}}{H\dot{H}}-\frac{2\dot{H}}{H^2}\approx
4  M_P^2  \left\{
\left[\frac{H'(\phi)}{H(\phi)}\right]^{2}
-\frac{H''(\phi)}{H(\phi)}
\right\}
 \, ,
\label{eq12bdd}
\end{eqnarray}
and the e-folding number is still given by (\ref{eq15}).

Let us now apply the above for the case of the power-law function of (\ref{powansatz}),
namely for $H(\phi)=\alpha \phi^n$. In this case  Eq. (\ref{phidot}) becomes
\begin{eqnarray}
\label{eq18nn}
&&\dot{\phi}\approx -2M_P^2\,n \alpha{\phi^{n-1}}\, ,
\end{eqnarray}
while  (\ref{Potential1}) reads
\begin{eqnarray}
\label{eq19nn}
&&\!\!\!\!\!\!\!
V(\phi)=\frac{6}{M_P^2} \alpha^{2} \phi^{2n}
-\frac{2}{M_P^4} \alpha^{2} {n^2}\phi^{2(n-1)} .
\end{eqnarray}
Additionally, the slow-roll parameters (\ref{eq12dd}),(\ref{eq12bdd}) become
\begin{eqnarray}
\label{eq12dd2}
&&\epsilon_1(\phi)
\approx
2  M_P^2 \frac{n^2}{\phi^2},
\\
&&
\epsilon_2(\phi)
\approx
4  M_P^2 \frac{n}{\phi^2},
\label{eq12bdd2}
\end{eqnarray}
while  the scalar field at the end of inflation, corresponding to
$\epsilon_1(\phi_e)=1$, is given by
\begin{eqnarray}
\label{eq20dd}
\phi_{e}=  \sqrt{2}nM_P.
\end{eqnarray}
Additionally, using (\ref{eq15})  and
(\ref{phidot}), the scalar field at the beginning of inflation becomes
\begin{equation}
\label{eq21nn}
{\phi _i} = M_P  \sqrt{ 2 n  (n+2N) }.
\end{equation}
Hence, inserting these in the expressions of the inflationary observables we finally
obtain
\begin{eqnarray}
\label{eq23nn}
&&{n_S} - 1 \simeq
-2\epsilon_1(\phi_{i})-2\epsilon_2(\phi_{i})
=- \frac{ 2 (n + 2)}{n+2N},
\\
\label{eq24nn}
&&
n_{T}\simeq-2\epsilon_1(\phi_{i})=- \frac{ 2n}{n+2N},\\
&&
r=-8n_T\, .
\label{eq23bbnn}
\end{eqnarray}

Similarly, for the   case of the exponential function of (\ref{expansatz1}),
namely for $H(\phi)=\beta e^{q\phi}$, we obtain
\begin{eqnarray}
\label{eq18nn22}
&&\dot{\phi}\approx -2M_P^2\,\beta q e^{q\phi}\, ,
\end{eqnarray}
while  (\ref{Potential1}) reads
\begin{eqnarray}
\label{eq19nn22}
&&\!\!\!\!\!\!\!
V(\phi)=\frac{6}{M_P^2} \beta ^{2}   e^{2q\phi}
-\frac{2}{M_P^4} \beta^2 q^2    e^{2q\phi}.
\end{eqnarray}
However, the slow-roll parameters (\ref{eq12dd}),(\ref{eq12bdd}) become
\begin{eqnarray}
\label{eq12dd2b}
&&\epsilon_1(\phi)
\approx
2  q^2 M_P^2 ,
\\
&&
\epsilon_2(\phi)
\approx
0.
\label{eq12bdd2b}
\end{eqnarray}
Thus, we can immediately see that the exponential form in the case of minimal coupling
cannot describe inflation successfully, since $\epsilon_1=const.$.

One can see that the above relations, which have been extracted in the case
of minimal-coupling, are different from the expressions of the previous subsections which
were extracted in the case of nonminimal derivative coupling. In particular, in
the  power-law   case relations (\ref{eq23nn})-(\ref{eq23bbnn}) are different
from  (\ref{eq23})-(\ref{eq23bb}), while in the exponential case a
successful realization of inflation is not possible. Furthermore,
relations (\ref{eq23nn})-(\ref{eq23bbnn}) cannot
fit the observational data. Actually, these disadvantages were one of the reasons that
inflation with nonminimal derivative coupling was introduced in \cite{Germani:2010gm}.
Hence, we conclude that the  nonminimal derivative coupling plays an important role, both
quantitatively (one has an additional parameter to fit the data) and qualitatively (the
unitarity issue is solved).

\section{Conclusion}
\label{Conc}

In this work we have investigated inflation with nonminimal derivative coupling through
the Hamilton-Jacobi formalism. In the majority of inflationary models one can perform the
conformal transformation to the Einstein frame, where the calculation of inflationary
observables is straightforward, however in  inflation with nonminimal derivative such a
transformation is absent and thus a detailed and lengthy perturbation analysis is needed
in order to provide inflationary observables that could be compared with the data. Hence,
in such a case the Hamilton-Jacobi analysis proves to be a very convenient method to
extract the scenario prediction for various observables.

In the Hamilton-Jacobi formalism one rewrites the cosmological equations by
considering the scalar field to be the time variable, which is always possible during the
slow-roll era, where the scalar varies monotonically (the extension to the
post-inflationary, oscillatory epoch is straightforward by matching together separate
monotonic epochs). We have performed such an analysis in inflation with nonminimal
derivative coupling, and we have applied it to two specific models for the Hubble
function, namely the power-law and the exponential cases. For each model we have
calculated various observables such as the tensor-to-scalar ratio, and the spectral index
and its running, and we have compared them with the Planck data.

For the case of power-law form we have shown that the tensor-to-scalar ratio and the
tensor spectral index have a linear dependence on the scalar spectral index, with its
properties
depending on the model parameters and the e-folding number, and confrontation with 2013
and 2015 Planck results shows a very good agreement. Additionally, the predictions for
the running spectral index are also in very good agreement with the data. Finally, for
transparency we have provided the corresponding profile of the scalar potential.

For the case of exponential form the tensor-to-scalar ratio and the tensor spectral index
have a
linear dependence on the scalar spectral index too, however not depending on the model
parameters or the e-folding number. Nevertheless, the predictions are in good agreement
with observational data. The behavior of the running spectral index is more complicated,
but still in very good agreement with the Planck results.

Finally, comparing our results with the case of minimal coupling, we saw that
the nonminimal derivative coupling can fit the data more efficiently, as well as solve
the unitarity problems of the minimal theory.

The above features reveal that the  Hamilton-Jacobi formalism can be very efficient in
analyzing inflationary models which do not allow for a conformal transformation to the
Einstein frame. And indeed, its application shows that inflation with nonminimal
derivative coupling is in agreement with observations and thus it can be a good candidate
for the description of Nature.\\

\section*{Acknowledgements}

HS would like to thank Iran's National Elites Foundation for financially
support during this work. He  expresses his appreciation to the Prof. Y. Sobouti for
sharing their pearls of wisdom with him during the course of this research. HS is also
immensely grateful to K. Karami and A. Mohammadi, for their comments on earlier versions
of the manuscript. This article is also based upon work from COST action CA15117
(CANTATA), supported by COST (European Cooperation in Science and Technology).

\end{document}